\documentstyle[12pt]{article}
\begin{document}
\begin{titlepage}
\hfill{PURD-TH-93-12}
\vskip 2cm
\centerline{\Large \bf VERTICES AND THE CJT EFFECTIVE POTENTIAL}
\vskip 1cm
\centerline{Bijan Haeri, Jr.}
\vskip 1cm
\centerline{\it Department of Physics}
\centerline{\it Purdue University}
\centerline{\it West Lafayette, Indiana 47907-1396}
\vskip 1.5cm
\centerline{ABSTRACT}
 \noindent
The Cornwall-Jackiw-Tomboulis effective potential is modified to include
a functional dependence on the fermion-gauge particle vertex, and
applied to a quark confining model of chiral symmetry breaking.
\end{titlepage}

\leftline{\bf 1. Introduction}
\vskip 5mm

The study of dynamical chiral symmetry breaking has benefited greatly
from the Cornwall-Jackiw-Tomboulis (CJT) effective potential \cite{cjt}.
The CJT potential is an effective potential of composite operators,
specifically
the propagators of a theory. Minimizing the CJT potential with respect to a
propagator gives the Schwinger-Dyson (SD) equation for that
propagator. The SD equation can have more than one solution. These solutions
must be inserted in the CJT potential to find the solution corresponding
to a stable minimum in the CJT potential, thus determining the physical
solution. In its simplest form the CJT potential as a functional
of propagators is of practical use only in solving propagator SD
equations, which have a severe truncation of their vertex SD equations,
as is the case for the ladder approximation, where the
full vertex is replaced by the bare vertex. The usual CJT potential does not
have a functional dependence on the vertices of a theory,
as a result the SD equations of the vertices cannot be found by minimizing the
CJT potential. There exist aternative means of constructing full vertices,
such as solving the Ward identity of the vertex (the gauge technique ), to
arrive at a full vertex as a functional of the propagators, which is
valid in the infrared regime. Putting the gauge technique vertex in the
propagtor SD equation by hand is inconsistent with minimizing
the CJT potential, which is not a functional of the gauge technique vertex.
Therefore testing the stability of the CJT potential with solutions found
from propagator SD equations containing the gauge technique
vertex can be unreliable (as will be demonstrated in section 3).

We will extend the CJT formalism to write an effective potential
for QED which has a functional dependence
on the fermion-fermion-gauge particle vertex in addition to the propagators,
and apply it to a confining model of dynamical chiral symmetry breaking.
Interestingly, it will be found that the chiral symmetry breaking solution
of the effective quark propagator SD equation minimizes the modified CJT
potential. Section 2 contains a derivation of the modified
CJT potential for QED, while section 3 applies the results to a
specific model.
\vfill
\eject

\leftline{\bf 2. CJT Potential as a Function of the Vertex}
\vskip 5mm

Consider massless QED with its gauge particle and fermion propagators, and
fermion-fermion-gauge particle vertex
denoted as $\Delta_{\mu \nu}(x,y)$, $S(x,y)$, and $\Gamma_{\mu}(x,y,z)$
respectively. The CJT effective action is given by
\begin{eqnarray}
\Gamma(\Delta_{\mu \nu},S)=\Gamma_{0}(\Delta_{\mu \nu},S)+
\Gamma_{2}(\Delta_{\mu \nu},S),
\end{eqnarray}
where
\begin{eqnarray}
\Gamma_{0}(\Delta_{\mu \nu},S)=
              &{{-i\hbar}\over{2}}Tr\ ln(
{{\Delta_{0}}_{\mu \alpha}}^{-1}\Delta_{\alpha \nu})
+{{i\hbar}\over{2}}Tr({{\Delta_{0}}_{\mu \alpha}}^{-1}
\Delta_{\alpha \nu}-g_{\mu \nu}) \nonumber \\
              &+i\hbar Tr\ ln(S_{0}^{-1}S)-i\hbar Tr(S_{0}^{-1}S-1),
\end{eqnarray}
while $\Gamma_{2}(\Delta_{\mu \nu},S)$ is the infinite sum over all possible
2 point-irreducible (2PI) Feynman graphs involving
$\Delta_{\mu \nu}$ and $S$,
and ${\Delta_{0}}_{\mu \nu}$ and $S_{0}$ are the bare gauge particle and
fermion propagators respectively.
Including a vertex source in the generator of connected Green's function, $W$,
gives
$W=W(G_{\mu \nu},K,\rho_{\mu})$, where $G_{\mu\nu}$, $K$, and $\rho_{\mu}$ are
the sources for $\Delta_{\mu \nu}$, $S$, and $\Gamma_{\mu}$ respectively.
Perfoming a triple Legendre transformation, we find
\begin{eqnarray}
\Gamma (\Delta_{\mu \nu},S,\Gamma_{\mu})=
 &W-i\hbar\int d^4xd^4y\left({1\over2}G_{\mu \nu}(x,y)\Delta^{\mu \nu}
(y,x)-K(x,y)S(y,x)\right)\nonumber \\
           &-i\hbar\int d^4xd^4yd^4z \rho^{\mu}(x,y,z)\Gamma_{\mu}(z,y,x),
\end{eqnarray}
corresponding to
\begin{eqnarray}
\Gamma(\Delta_{\mu \nu},S,\Gamma_{\mu})=\Gamma_{0}(\Delta_{\mu \nu},S)+
\Gamma_{2}(\Delta_{\mu \nu},S,\Gamma_{\mu}),
\end{eqnarray}
where $\Gamma_{2}(\Delta_{\mu \nu},S,\Gamma_{\mu})$ is an infinite sum over
all 2PI Feynman graphs containing all possible interactions of full propagators
and full vertices, as well as all the Feynman graphs that make up
 $\Gamma_{2}(\Delta_{\mu \nu},S)$. Evaluating the effective action at a
stationary point gives
\begin{eqnarray}
{{\delta \Gamma}\over{\delta
\Delta_{\mu \nu}}}=-{{i\hbar}\over{2}}G_{\mu \nu}|_{G_{\mu \nu}=0},
\end{eqnarray}
\begin{eqnarray}
{{\delta \Gamma}\over{\delta S}}=i\hbar K|_{S=0},
\end{eqnarray}
and
\begin{eqnarray}
{{\delta \Gamma}\over{\delta \Gamma_{\mu}}}=-i\hbar
\rho_{\mu}\vert_{\rho_{\mu}=0},
\end{eqnarray}
where Eq.'s\ 5-7 are the
SD equations for $\Delta_{\mu \nu}(x-y)$, $S(x-y)$, and $\Gamma_{\mu}(x-y,y-z)$
respectively, where the propagators(vertex) are(is) a funtion of 1(2)
variables,
rather than 2(3) before the sources were set to zero.
At the stationary point the effective action can be written as
\begin{eqnarray}
\Gamma(\Delta_{\mu \nu},S,\Gamma_{\mu})=-V(\Delta_{\mu \nu},S,\Gamma_{\mu})
\int d^4x,
\end{eqnarray}
where the effective potential, $V(\Delta_{\mu \nu},S,\Gamma_{\mu})$, is given
by \cite{cnwl}
\begin{eqnarray}
V(\Delta_{\mu \nu},S,\Gamma_{\mu})=V_0(\Delta_{\mu \nu},S,\Gamma_{\mu})+
V_{2}(\Delta_{\mu \nu},S,\Gamma_{\mu}),
\end{eqnarray}
\begin{eqnarray}
V_0(\Delta_{\mu \nu},S,\Gamma_{\mu})=&-i\hbar\int {{d^4p}\over{2{\pi}^4}}
Tr\Biggl(-{1\over2}ln({{\Delta_{0}}_{\mu\alpha}}^{-1}(p)
\Delta_{\alpha \nu}(p)) \nonumber \\
&+{1\over{2}}({{\Delta_{0}}_{\mu \alpha}(p)}^{-1}
\Delta_{\alpha \nu}(p)-g_{\mu \nu}) \nonumber \\
             &+ln(S_{0}(p)^{-1}S(p))-(S_{0}(p)^{-1}S(p)-1)\Biggr),
\end{eqnarray}
where $V_{2}(\Delta_{\mu \nu},S,\Gamma_{\mu})$ is the infinite sum over
vacuum graphs shown in Fig. 1.
Eqs. 5-7 can alternatively be written as
\begin{eqnarray}
{{\delta V}\over{\delta
\Delta_{\mu \nu}(p)}}=0,
\end{eqnarray}
\begin{eqnarray}
{\delta V\over{\delta S(p)}}=0,
\end{eqnarray}
and
\begin{eqnarray}
{\delta V\over{\delta \Gamma_{\mu}(p,q)}}=0,
\end{eqnarray}
which are displayed in Figs. 2-4 respectively.
The coffecients of the graphs on the right hand side of Fig. 1 are
determined by first examining
the vertex SD equation shown in Fig. 4, and then either of the
propagator SD equations (Figs.\ 2 and 3); using Fig.\ 4 it can be seen
that $V_{2}(S,\Gamma_{\mu})=V_{2}(S)$ as we expect.

The vertex SD equation is not solvable as the infinite series shown in
Fig.\ 4. Truncating the series beyond the first graph on the right hand
side of Fig.\ 4, and keeping only the first graph on the right hand side
of Fig.\ 2 (choosing the Landau gauge) is
a common approximation (ladder approximation) used to solve the fermion SD
equation. Another approximation scheme  of determining the
vertex known as the modified gauge technique is to formulate
an ansatz for the vertex in terms of the fermion propagator
that solves the Ward identity for the vertex and preserves
the multiplicative renormalizability and gauge covariance of the
fermion SD equation \cite{jk,bh2}.
But since the vertex ansatz depends on the fermion propagator, the fermion
SD equation is modified to give

\begin{eqnarray}
 {{\delta V}\over{\delta S}}={{\delta V}\over{\delta S}}\Biggr\vert_
{\Gamma_{\mu}}+{{\delta \Gamma_{\mu}}\over{\delta S}} {{\delta V}
\over{\delta \Gamma_{\mu}}}\Biggr\vert_{S}=0.
\end{eqnarray}
The first term in the middle corresponds to the usual
 fermion SD equation (Eq.\ 12) when the second term in the middle vanishes,
that is the vertex SD equation
(Eq.\ 13) is satisfied, otherwise there will be extra
terms in the fermion SD equation (besides those appearing on the right hand
side of Fig. 3).

\vskip 1cm
\leftline{\bf 3. Application to Models of Confinement}
\vskip 1cm

Quark confinement in QCD has long been modeled by replacing the gluon
propagator
appearing in the effective quark propagator
SD equation with a $1\over{k^4}$ potential, representing
a linearly rising potential in coordinate space \cite{cc}.
The $1\over{k^4}$ potential has an
infrared divergence, which is commonly regulated to give a
$\delta^4(k)$ potential \cite{pagels}. Here we will follow this example and
replace the gluon propagator with:
\begin{eqnarray}
\Delta_{\mu \nu}(k)=ag_{\mu \nu}{{16\pi^4}\over{C_{f}}}\delta^4(k),
\end{eqnarray}
where $C_f={{N^2-1}\over{2N}}$ is the Casimir eigenvalue
of the of the fundamental representation
of $SU(N)$, and $\sqrt{a}$ is a scale associated with confinement.
It is usual at this point to completely truncate the quark-quark-gluon (qqg)
vertex
by replacing the full qqg vertex $\Gamma_{\mu}$, by a bare vertex
$\gamma_{\mu}$ in the quark propagator
SD equation, and thus leaving one SD equation to solve \cite{mn}.
We will alternatively solve the quark propagator SD equation with the full
qqg vertex intact. The actual qqg vertex we will use is the modified qqg
vertex found by resumming Feynman diagrams to redefine the n-point Green's
functions in terms of modified Green's functions \cite{bh1,hh}.
In this case the modified gluon propagator $\hat{\Delta}_{\mu \nu}(k)$
is replaced by Eq. 15,
while the full modified qqg vertex, $\hat{\Gamma}_{\mu}(p,k)$ satisfies a
simple Ward identity
\begin{eqnarray}
  \hat{\Gamma}_{\mu}(p,p)={{\partial S^{-1}(p)}\over{\partial p_{\mu}}}.
\end{eqnarray}
Using Eq.\ 16, $\hat{\Gamma}_{\mu}(p,p)$ can be solved exactly in terms of
$S(p)$ without the need to solve the vertex SD equation of Fig. 4.
Writing $S(p)$ in terms of scalar functions
\begin{eqnarray}
 S(p)={1\over{\alpha \rlap /p -\beta (p)}},
\end{eqnarray}
we find that
\begin{eqnarray}
 \hat{\Gamma}_{\mu}(p,p)=\alpha(p)\gamma_{\mu}+
2p_{\mu}\rlap /p {{\partial \alpha(p)}\over{\partial p^2}}-
\{ \gamma_{\mu},\rlap /p\} {{\partial \beta(p)}\over{\partial p^2}}
\end{eqnarray}
satisfies Eq.\ 16.
It is useful to multiply Eq.\ 18 on the left and right by $S(p)$, so that
it becomes
\begin{eqnarray}
  S(p)\hat{\Gamma}_{\mu}(p,p)S(p)=-{{\partial S(p)}\over{\partial p_{\mu}}},
\end{eqnarray}
and write $S(p)$ as
\begin{eqnarray}
 S(p)=A(p)\rlap /p+B(p),
\end{eqnarray}
giving
\begin{eqnarray}
 S(p)\hat{\Gamma}_{\mu}(p,p)S(p)=-A(p)\gamma_{\mu}-
2p_{\mu}\rlap /p {{\partial A(p)}\over{\partial p^2}}-
\{ \gamma_{\mu},\rlap /p\} {{\partial B(p)}\over{\partial p^2}},
\end{eqnarray}
where  $\alpha(p)$ and $\beta(p)$ are related to $A(p)$ and $B(p)$ by
\begin{eqnarray}
\alpha(p)={{A(p)}\over{p^2A^2(p)-B^2(p)}}
\end{eqnarray}
\begin{eqnarray}
\beta(p)={{B(p)}\over{p^2A^2(p)-B^2(p)}}
\end{eqnarray}
The SD equation for the effective quark propagator is
\begin{eqnarray}
S(p)=S_0(p)+ig^2C_fS_0(p)\int {{d^4k}\over{16\pi^4}} \gamma_{\mu}S(k)
\hat{\Gamma}_{\nu}(k,p)S(p)\hat{\Delta}^{\mu \nu}(p-k),
\end{eqnarray}
where
\begin{eqnarray}
S_0(p)={1\over{\rlap /p}}.
\end{eqnarray}
Substituting Eq.\ 15, and Eq.\ 21 into Eq.\ 24, we find
\begin{eqnarray}
p^2{{\partial A(p)}\over{\partial p^2}}=
{1\over{2a}}\left((p^2-4a)A(p)-1\right)
\end{eqnarray}
\begin{equation}
{{\partial B(p)}\over{\partial p^2}}={{B(p)}\over{2a}}
\end{equation}
Eqs. 26,\ 27,\ 22,\ and 23 have been solved in Euclidean space
(i.e. $p^2=-{p_E}^2$) by Burden, Roberts, and Williams \cite{brw}, who find
for the case of chiral symmetry breaking,
\begin{equation}
A(x)=-{1\over{2a}}{{x-1+e^{-x}}\over{x^2}},
\end{equation}
\begin{equation}
B(x)={1\over{2\sqrt{a}}}e^{-x},
\end{equation}
\begin{equation}
\alpha_b(x)={{2x(e^{-x}+x-1)}\over{x^3e^{-2x}+2(e^{-x}+x-1)^2}},
\end{equation}
\begin{equation}
\beta(x)=-\sqrt{a}{{2x^3}\over{x^3e^{-x}+2e^x(e^{-x}+x-1)^2}},
\end{equation}
where $x={{{p_E}^2}\over{2a}}$. For the chiral symmetry preserving
case $A(x)$ has the same solution as in Eq.\ 28, while
\begin{equation}
\alpha_s(x)={{x}\over{e^{-x}+x-1}},
\end{equation}
\begin{equation}
\beta(x)=B(x)=0.
\end{equation}
$\alpha_s(x)$, and  $\alpha_b(x)$ are the solutions of $\alpha(x)$
corresponding to the chiral symmetric and chiral symmetry breaking cases
respectively.
To determine which solution is prefered, we need to evalute both sets of
solutions in Eq.\ 9, namely we must calculate
\begin{eqnarray}
\Delta V\equiv V_{s}-V_{b},
\end{eqnarray}
where $V_{s}$ is the modified CJT effective potential evaluated with the
chiral symmetric solution to $S(p)$, and $\Gamma_{\mu}(p,p)$ (Eq.\ 28, 32,
and 33), while $V_{b}$
 uses the chiral symmetry
breaking solution (Eq.\ 28-31).
Defining
\begin{equation}
\tilde{V}\equiv {1\over{n_fNa^2}}V(S,\Gamma_{\mu}),
\end{equation}
where $n_f$ is the number of fermions, we find as in Ref.\ 10 that,
\begin{equation}
\tilde{V_0}_{s}-\tilde{V_0}_{b}=-{1\over{2\pi^2}}\int xdx\
ln\Biggl(1+{{\bar{B}^2}\over{2x\bar{A}^2}}\Biggr)=-.04,
\end{equation}
where $\bar{A}\equiv aA$, and $\bar{B}\equiv aB$.
$V_2(S)$ in Fig.\ 1b does not contribute to Eq.\ 34.
The second and third terms in the series on the right hand side of Fig.\ 1a
have
contributions to Eq.\ 34 given by,
\begin{eqnarray}
\Delta\tilde{V}|_{2}=&-{1\over{\pi^2}}\int xdx
\Biggl(x(\bar{A}+2x\bar{A}')(\alpha_s'-\alpha_b')
+(2\bar{A}+x\bar{A}')
(\alpha_s-\alpha_b) \nonumber \\
&-x\beta'B'\Biggl),
\end{eqnarray}
and
\begin{eqnarray}
\Delta\tilde{V}|_{3}=&-{2\over{\pi^2}}\int xdx
\Biggl([x\bar{A}^2-2x^2\bar{A}'(\bar{A}+x\bar{A}')]
(\alpha_s\alpha_s'-\alpha_b\alpha_b'
+x(\alpha_s'^2-\alpha_b'^2)) \nonumber \\
&+(\bar{A}^2+x \bar{A}\bar{A}'+x^2\bar{A}'^2)
(\alpha_s^2-\alpha_b^2-x\bar{\beta}^2) \nonumber \\
&+x({1\over{2}}\alpha_b \bar{A}+x\alpha_b
\bar{A}'+x\alpha_b' \bar{A}
+2x^2\alpha_b'\bar{A}')\beta'B'\nonumber \\
&-x(\alpha_b^2+x\alpha_b\alpha_b'+x^2\alpha_
b'^2)\bar{B}'^2+{1\over 2}x^2\beta'^2B'^2\Biggr),
\end{eqnarray}
where $\bar{\beta}\equiv a\beta$, and prime denotes partial deifferentiation
with
respect to x.
Eq.\ 38 has an infrared singularity in
 $\tilde{V_2}_{s}|_{3}$, which originates from a mismatch in
infrared singularities between the left and right sides of Fig.\ 4.
As x approaches zero we find that $\alpha_S={2\over{\epsilon}}$, where
$\epsilon$ is infinitesmal; therefore in the chiral symmetric case, Fig.\ 4
will have a singularity structure that goes as
\begin{equation}
{b\over{\epsilon}}={c\over{\epsilon^2}}+{d\over{\epsilon^3}}+...,
\end{equation}
where b, c, and d are constants. Presumably the right hand side will
sum up to give a power of $1\over\epsilon$, which can be achieved by
replacing $\delta^4(k)$ in Eq.\ 15 with $\epsilon\delta^4(k)$
in the limit of small x to obtain a power of $1\over\epsilon$ for each term on
the right hand side of Fig.\ 4. But such a resummation is only valid in the
infrared regime, and we are interested in performing an integral over all
momenta. We notice instead that the infrared singularity in Eq.\ 38 is isolated
to the term involving $\alpha_{s}^2\bar{A}^2$ (other terms are also
infrared singular,
but the singularity cancels among them); in fact in graphs involving more
than four vertices the singularity is also dominate in
the term containing a power
of $\alpha_{s}\bar{A}$. We find that terms invoving $\alpha_{s}\bar{A}$ and
$\alpha_{b}\bar{A}$ resum
into a logarithm for the graphs shown in Fig.\ 1a (and the graph involving 6
vertices, we have not checked 8 and higher vertex graphs).
Replacing the sum of the terms on the right hand sides of Eqs.\ 37
and \ 38 involving $\bar{A}(\alpha_{s}-\alpha_{b})$, and
${\bar{A}}^2({\alpha_{s}}^2-{\alpha_{b}}^2)$ respectively by,
\begin{eqnarray}
-{1\over{\pi^2}}\int xdx\Biggl(4(\alpha_{s}-\alpha_{b})\bar{A}
+ln\Biggl({{1-2\alpha_{s}\bar{A}}\over{1-2\alpha_{b}\bar{A}}}\Biggr)\Biggr),
\end{eqnarray}
and then adding Eqs.\ 36, 37, and 38, we find that
\begin{eqnarray}
{\tilde{V}_{s}}-{\tilde{V}_{b}}=.1>0.
\end{eqnarray}
To accurately determine ${\tilde{V}_{s}}-{\tilde{V}_{b}}$, we need
to sum the infinite terms on the right hand side of Eq.\ 1a for both
the symmetric and symmetry breaking cases, which may be tractable, but
will not be pursued here.
By keeping 3 terms in the series
in Fig.\ 1a, we have demonstrated the possibility that the
chiral symmetry breaking
solution is the physical solution to Eq.\ 24 unlike
the conclusion reached
in Ref.\ 11.
\vfill
\eject
\leftline{\bf Acknowledgements}
\vskip 1cm
\noindent
The author is grateful for helpful discussions with J.M. Cornwall.
The author also benefited from discussions with S.T. Love and C.D.
Roberts.
This work was supported by the US Department of Energy under contract
DE-AC02-76ER01428 (Task B).

\vskip 1cm
\leftline{\bf Figure Captions}
\vskip 1cm
\noindent
{\bf Figure 1.} The interaction part of the effective potential.

\noindent
{\bf Figure 2.} The gauge particle propagator SD equation.

\noindent
{\bf Figure 3.} The fermion propagator SD equation.

\noindent
{\bf Figure 4.} The vertex SD equation.
\vfill
\eject

\end{document}